\documentclass[prc,aps,preprint,superscriptaddress,showpacs]{revtex4}
\usepackage{epsfig}
\usepackage{graphicx}
\usepackage{amsmath}
\usepackage{yhmath}
\usepackage{amssymb}
\usepackage{bm}

\date{\today}

\tighten

\newcommand{\disregard}[1]{}

\newcommand{\be}{\begin{equation}}
\newcommand{\ee}{\end{equation}}
\newcommand{\bn}{\begin{eqnarray}}
\newcommand{\en}{\end{eqnarray}}
\newcommand{\ba}{\begin{array}}
\newcommand{\ea}{\end{array}}
\newcommand{\bc}{\begin{center}}
\newcommand{\ec}{\end{center}}
\newcommand{\bml}{\begin{mathletters}}
\newcommand{\eml}{\end{mathletters}}

\newcommand{\etal}{{\it et al.}}

\newcommand{\ls}{{$\ell s$}}
\newcommand{\shf}{{{\sc shf}}}
\newcommand{\todd}{{{\sc to}}}

\tighten

\begin{document}

\draft

\preprint{\fbox{\sc version of \today}}


\title{Probing the nuclear energy functional at band
termination}

\author{{\sc  Honorata Zdu\'nczuk}}
\email{             zdunczuk@fuw.edu.pl}
\affiliation{Institute of Theoretical Physics, University of Warsaw,\\
             ul. Ho{\.z}a 69, PL-00 681 Warsaw, Poland}

\author{{\sc  Wojciech Satu{\l}a}}
\email{             satula@fuw.edu.pl}
\affiliation{Institute of Theoretical Physics, University of Warsaw,\\
             ul. Ho{\.z}a 69, PL-00 681 Warsaw, Poland}
\affiliation{          KTH (Royal Institute of Technology),\\
           AlbaNova University Center, 106 91 Stockholm, Sweden}

\author{{\sc Ramon A. Wyss}}
\email{           wyss@kth.se}
\affiliation{          KTH (Royal Institute of Technology),\\
           AlbaNova University Center, 106 91 Stockholm, Sweden}

\begin{abstract}
A systematic study of terminating states in $A$$\sim$50
mass region using the self-consistent Skyrme-Hartree-Fock
model is presented.
The objective is to demonstrate that
the terminating states, due to their intrinsic simplicity, offer
unique and so far unexplored opportunities to study different
aspects of the effective NN interaction
or nuclear local energy density functional. In particular, we
demonstrate that the agreement of the calculations to the data
depend on the spin fields and the spin-orbit term  which, in turn, allows to
constrain the appropriate Landau parameters and the
strength of the spin-orbit potential.
\end{abstract}


\maketitle

\newpage


\section{Introduction}

Within the mean-field approach the
spin-orbit (\ls) splitting is usually studied
via the comparison of theoretical and experimental
single-particle ($sp$) energies of the \ls-doublet. The
requirement is that both \ls-partners should be
simultaneously occupied~\cite{[Rut98],[Lop00]}. The method assumes that
under such conditions the  core polarization effects, which are known to modify
strongly {\it sp\/} energies~\cite{[Ham76],[Ber80]}, are similar for
\ls-partners and therefore do not affect the
\ls-splittings, at least not in a major way.

The method requires by definition precise empirical knowledge on
$sp$ energies of deep-hole states which are very difficult to measure.
In addition, particle vibration coupling may contribute to the
splitting and perturb the pure $sp$ picture.
The available data on \ls-splittings and their isotopic or
isotonic dependence are therefore both scarce and uncertain.
For example, in the $A$$\sim$40 mass region, which is of primary interest
in this work, the most recent
evaluations~\cite{[Oro96]} give
$\Delta\varepsilon_{d_{3/2}-d_{5/2}}$$\approx$\,6\,MeV  in $^{40}$Ca and
$\Delta\varepsilon_{d_{3/2}-d_{5/2}}$$\approx$\,5\,MeV  in $^{48}$Ca,
respectively, while older works give
$\Delta\varepsilon_{d_{3/2}-d_{5/2}}$$\approx$\,6.8\,MeV~\cite{[Swi66]},
$\approx$\,7.3\,MeV~\cite{[Tyr66]}, and  $\approx$\,7.7\,MeV~\cite{[Ray79]} in
$^{40}$Ca and $\approx$\,5.3\,MeV~\cite{[Ray79]}  in  $^{48}$Ca.
More detailed information on $sp$ levels can be found in Ref.~\cite{[Isa02]}.

In this work we want to pursue a novel method to study
the \ls-potential by exploring high-spin data. The method is based on
a direct comparison of the excitation
energies of terminating states [which are maximum-spin states
within a given $sp$ configuration] for two
carefully selected configurations.  In this study
we limit ourselves to  the $d_{3/2}^{-1}f_{7/2}^{n+1}$
and $f_{7/2}^n$ configurations in
$A$$\sim$50, 20$\leqslant$Z$<$N$\leqslant$24 nuclei, where
$n$ denotes the number of valence particles outside
the $^{40}$Ca core.
The difference, $\Delta E$, between the excitation
energies of states terminating within the $d_{3/2}^{-1}f_{7/2}^{n+1}$ and
$f_{7/2}^n$ configurations is dominated by the size of the
magic gap 20. The magnitude of the magic gap 20
is in turn governed by the strength of the \ls-potential.
Indeed, for the spherical Nilsson Hamiltonian~\cite{[Nil55]},
i.e. the three-dimensional harmonic-oscillator (HO) potential augmented by a
one-body spin-orbit $ - 2\kappa \hbar\omega_o{\boldsymbol \ell}{\boldsymbol s}$
and orbit-orbit $-  \kappa \mu \hbar\omega_o {\boldsymbol \ell}^{\, 2}$ term
one has:
\be\label{nilson}
   \hat H_{Nilsson}
       - \frac{3}{2} \hbar\omega_o = \hbar\omega_o \left\{ N - \kappa \left[
       2 {\boldsymbol \ell} {\boldsymbol s}  + \mu ( {\boldsymbol \ell}^{\, 2}
            - \langle {\boldsymbol \ell}^{\, 2} \rangle_N ) \right] \right\}.
\ee
Hence, within the Nilsson model, which is considered to be a fundamental
approximation for the nuclear mean-field potential, the
magnitude of the magic gap 20, or more precisely the $f_{7/2}$$-$$d_{3/2}$
splitting reads:
\be\label{gap20}
  \Delta e_{20} = \hbar\omega_o  ( 1 - 6\kappa - 2\kappa\mu ).
\ee
In light nuclei the nuclear potential resembles the pure
HO, thus  $\mu \sim 0$.  Hence, within the Nilsson model, the \ls-term
plays a dominant role in establishing the size of the magic gap 20.
Indeed, $\hbar\omega_o$ determines the global energy scale in
low energy nuclear physics and one does not expect this value
to change substantially.
Although Eq.~(\ref{gap20}) pertains to the $f_{7/2}$$-$$d_{3/2}$
splitting, the conclusion drawn above
seems to be easily extendable to heavier nuclei
since there  $\mu$$\rightarrow$1/2 as a consequence of  approximate
pseudo-SU(3) symmetry~\cite{[Boh82],[Bah92],[Gin97]}.

By limiting ourselves to the study of band terminating states
only, we  access the regime of essentially unperturbed $sp$ motion,
where correlations going beyond mean-field are expected to be strongly
suppressed.
Indeed, the success of simple Nilsson-Strutinsky calculations of terminating
bands by Ragnarsson and coworkers, for review see~\cite{[Afa99]}, nicely
confirm the structural purity and $sp$ nature of the terminating states.
However, within the self-consistent approaches, in particular within the
Skyrme-Hartree-Fock (\shf) approach which is used here, there are additional
difficulties and in turn uncertainties related to the limited knowledge of
the time-odd components of the mean-field. Direct studies of these terms are
not only scarce but also in many cases inconclusive, see~\cite{[Sat04]} and
refs. quoted therein. In contrast, the structural simplicity of terminating
states is appealing and allows for a direct study of these terms.

The method proposed here to determine the $\ell s$-term has clear
advantages over the standard method mentioned at the
beginning: ({\it i\/}) it uses
terminating states which are probably {\it the best\/} examples of
unperturbed single-particle motion; ({\it ii\/})
the terminating states are uniquely defined implying that configuration
mixing going beyond mean-field is expected to be marginal;
({\it iii\/}) shape polarization is
included automatically within the calculations scheme
and no further {\it ad hoc\/} assumptions are necessary;
({\it iv\/}) a  wide set of rather precise experimental data
is already available throughout the periodic table.

The paper is organized as follows. Sect.~\ref{data}
overviews available empirical data. Sect.~\ref{edf}
briefly recalls the local energy density functional ({\sc ledf})
formulation  of the {\sc shf} method.
Sect.~\ref{sfield} reveals the problems related to spin fields
emerging from the Skyrme force induced local energy
density functional ({\sc s-ledf}) in $N$$\sim$$Z$ nuclei
both in the ground state as well as at the band termination.
It is shown, that an unification of spin fields
cures these problems leading to an unified
description of terminating states.  The remaining discrepancy
between experiment and theory can further be reduced by
tuning (weakening) the strength of
the spin-orbit interaction as shown in Sect.~\ref{so}. Finally, in
Sect.~\ref{addit} we test the additivity relation for the only known case
of a $2p$-$2h$ excitation across the magic gap
in $^{45}$Sc. All \shf~calculations presented in this paper
were done using the \shf~code
{\sc hfodd} of Dobaczewski, Dudek, and Olbratowski~\cite{[Dob00],[Dob04]}.



\section{Experimental data}\label{data}

\begin{table}
\begin{center}
\begin{tabular}{lccrrcrrr}
\hline
\hline
     &  Ref.  &  $\quad \quad f_{7/2}^n:$ & $E[I_{max}] $  & $I_{max}$  &
     $\quad \quad d_{3/2}^{-1}f_{7/2}^{n+1}:$  &   $E [I_{max}] $  &
   $I_{max}$ & $\quad \quad \quad \Delta E_{exp}$ \\
\hline
\hline
 $^{42}_{20}\mbox{Ca}_{22}$    &  \protect{\cite{[Lac03]}} &   &
          3.189 &  6$^+$   &  & 8.297  &  11$^-$ &  5.108   \\
\hline
 $^{44}_{20}\mbox{Ca}_{24}$    &  \protect{\cite{[Lenzi]}} &   &
         10.568 &  8$^+$    &  &  5.088  &   13$^-$     &   5.480    \\
\hline
 $^{44}_{21}\mbox{Sc}_{23}$    &  \protect{\cite{[Lac03a]}} &   &
          9.141   &  11$^+$    & &    3.567  &  15$^-$     &  5.574        \\
\hline
 $^{45}_{21}\mbox{Sc}_{24}$    &  \protect{\cite{[Bed01]}} &   &
           5.417   &  23/2$^-$    & &   11.022  &  31/2$^+$    &  5.605 \\
\hline
                               &  \protect{\cite{[Lenzi]}} &    &
             &      &  &  15.701  &  35/2$^-$  &  10.284  \\
\hline
 $^{45}_{22}\mbox{Ti}_{23}$    &  \protect{\cite{[Lenzi]}} &  &
        7.143    &  27/2$^-$   & &  13.028    &  33/2$^+$     &   5.885  \\
\hline
 $^{46}_{22}\mbox{Ti}_{24}$    &  \protect{\cite{[Lenzi]}} & &
         10.034   &  14$^+$   & &  15.549  &  17$^-$     &   5.515       \\
\hline
 $^{47}_{23}\mbox{V}_{24}$    &  \protect{\cite{[Bra01]}} & &
         10.004  &  31/2$^-$  & &   15.259  &  35/2$^+$   &  5.255       \\
\hline
\hline
\end{tabular}
\caption[A]{Spins and  excitation energies of terminating states in
$20\leqslant Z < N \leqslant 24$ nuclei. The first two columns
are representative the for $f_{7/2}^n$ configuration where $n$ denotes
the number of valence particles outside the $^{40}$Ca core.
The next two columns are representative for
the $d_{3/2}^{-1}f_{7/2}^{n+1}$ configuration
involving the $1p$-$1h$ proton excitation across the magic gap 20. For the
case of $^{45}$Sc, data on the $d_{3/2}^{-2}f_{7/2}^{n+2}$
configuration involving two
$1p$-$1h$ excitations from the $d_{3/2}$ to $f_{7/2}$ are also included.
The relative excitation energies between
the two configurations are given in the last column.
}
\label{edata}
\end{center}
\end{table}


All available experimental data on the terminating states for
$f_{7/2}^n$ and $d_{3/2}^{-1}f_{7/2}^{n+1}$ configurations in
$A$$\sim$50, 20$\leqslant$Z$<$N$\leqslant$24 nuclei where
both states are known are listed in Tab.~\ref{edata}.
In the present data set we have excluded the $N$=$Z$ nuclei since in
these nuclei the terminating state
$d_{3/2}^{-1}f_{7/2}^{n+1}$ is not uniquely defined.
Indeed, two states with the same value of $I_{max}$ can
be formed, having approximately
(due to the weak isospin symmetry breaking) symmetric
($T=0$) and antisymmetric ($T=1$) proton-neutron
configurations, respectively.
$N$=$Z$ nuclei will be addressed in a forthcoming paper~\cite{[Sat04b]}.

The differences in excitation energies, $\Delta E_{exp}$, listed
in the last column of Tab.~\ref{edata} are fairly constant.
The mean value equals $\overline{\Delta E_{exp}} = 5.489$\,MeV while the
standard deviation is $\sigma = 0.251$\,MeV  i.e. at the level of $\sim$\,5\%
only. This suggests that the bulk part of $\Delta E_{exp}$ is indeed related
to the energy of $1p$-$1h$ excitation across the gap 20, and that polarization
effects are either weak or, most likely, cancel out.

This conclusion is further supported by a recent measurement in $^{45}$Sc
of a terminating state at  spin $I^\pi_{max} = 35/2^-$ and
excitation energy of 15.701\,MeV~\cite{[Lenzi]}.
This state involves the $2p$-$2h$ excitation from
the $\pi d_{3/2}$ to $\pi f_{7/2}$
sub-shell. The excitation energy with respect to the aligned
$f_{7/2}^n$ state equals $\Delta E_{2p2h}=10.284$\,MeV.
For the extreme case of pure additivity this excitation can be composed
from two, uniquely defined, $1p$-$1h$ states terminating at
$I^\pi_{max} = 29/2^+$ (termination at unfavored signature)
which both are known.
The empirical excitation energies of these states, relative
to the aligned $f_{7/2}^n$ state, are
4.753\,MeV and 5.783\,MeV, respectively. The resulting sum yields
$\Delta E^{(add)}_{2p2h}$=10.536\,MeV implying that
additivity holds within $\sim$2\%.

\section{Skyrme-Hartree-Fock local energy density functional}\label{edf}

The starting point of the \shf~approach is an energy density functional
which, in the isoscalar-isovector
$t=0,1$ representation, takes the following form:
\begin{equation}\label{efun}
{\cal E}^{Skyrme}=\sum_{t=0,1}\int
  d^3{\boldsymbol r}
\left[ {\mathcal H}_t^{(TE)} ({\boldsymbol r}) +
       {\mathcal H}_t^{(TO)} ({\boldsymbol r}) \right].
\end{equation}
The local energy density functional ${\mathcal H}$ ({\sc ledf}) is uniquely
expressed as a bilinear form of time-even ({\sc te}) $\rho, \tau,
\overleftrightarrow{J}$  and time-odd ({\sc to}) $\boldsymbol{s},
\boldsymbol{T}, \boldsymbol{j}$ local densities, currents, and by their
derivatives: \begin{equation}\label{teven} {\mathcal
H}_{t}^{(TE)} ({\boldsymbol r}) = C_{t}^{\rho} \rho_t^2
+ C_{t}^{\Delta\rho} \rho_t\Delta\rho_t + C_t^\tau \rho_t\tau_t + C_t^J
\overleftrightarrow{J}_t^2 + C_t^{\nabla J} \rho_t
{\boldsymbol \nabla}\cdot {\boldsymbol J}_t,
\end{equation}
\begin{equation}\label{todd}
{\mathcal H}_{t}^{(TO)}
({\boldsymbol r}) = C_{t}^{s}
\boldsymbol{s}_t^2 + C_{t}^{\Delta s} \boldsymbol{s}_t\Delta \boldsymbol{s}_t
+ C_t^T \boldsymbol{s}_t\cdot \boldsymbol{T}_t + C_t^j {\boldsymbol j}_t^2 +
C_t^{\nabla j} \boldsymbol{s}_t  \cdot ( {\boldsymbol \nabla}\times
{\boldsymbol j}_t).
\end{equation}
The division of {\sc ledf} into {\sc te} and
{\sc to} parts is very convenient since the
latter contributes only when time reversal symmetry is broken.

In the above formulas $\overleftrightarrow{J}_t^2 \equiv \sum_{\mu\nu}
J_{\mu\nu,t}^2$ while the vector spin-orbit density,
${\boldsymbol J}_t$,  is not an independent
quantity but constitutes an antisymmetric part of the tensor
density, i.e.,
${\boldsymbol J}_t \equiv \sum_{\mu\nu} \epsilon_{\mu\nu} J_{\mu\nu,t}$.
Definitions of all local
densities and currents $\rho, \tau, \overleftrightarrow{J}, \boldsymbol{s},
\boldsymbol{T}, \boldsymbol{j}$ can be found in numerous references
and will not be repeated here.
We follow the notation used in~Refs.~\cite{[Dob95],[Dob00],[Ben02]} where
references to earlier works can be found as well.

By taking an expectation value of the Skyrme force over the Slater
determinant  one obtains  {\sc ledf}  (\ref{efun})--(\ref{todd})
with  20 coupling constants $C$ that are expressed uniquely through the 10
parameters $x_i, t_i, \, i=0,1,2,3$ and $W, \alpha$  of the standard
Skyrme force. The appropriate formulas can be found,
e.g., in Refs.~\cite{[Dob95],[Ben02]}.
Due to the local gauge invariance (which includes the Galilean
invariance) of the Skyrme force~\cite{[Eng75],[Dob95]}
only 14 coupling constants $C$ are independent quantities.
The local gauge invariance links three pairs of
time-even and time-odd constants in the following way:
\be\label{gauge}
C_t^j
= - C_t^{\tau},  \quad  C_t^J  = - C_t^T \quad C_t^{\nabla j}  = C_t^{\nabla
J}.
\ee

Since the \shf~approach uses interaction-induced
coupling constants $C$ it constitutes a
restricted version of the local energy density theory of
Hohenberg-Kohn-Sham~\cite{[Hoh64],[Koh65],[Koh98]} type.
However, only very few \shf~approaches rigidly enforce
the Skyrme-force-related values of $C$. Among those studied here
these include SkP~\cite{[Dob84]},
SkXc~\cite{[Bro98]}, and Sly5~\cite{[Cha97]}. Other forces
studied here,  including Sly4~\cite{[Cha97]}, SIII~\cite{[Bei75]},
SkO~\cite{[Rei99]}, and SkM$^\star$~\cite{[Bar82]}
disregard the tensor terms by setting $C_t^J = C_t^T\equiv 0$.
This is not only due to practical reasons (these terms are the most
difficult technically) but also due  to lack of clear experimental information
that would allow for reasonable estimates of their strengths.
Moreover, in the case of SkO we were forced to set $C_t^{\Delta s}\equiv 0$
to assure convergence. All versions of
{\sc ledf} that use the Skyrme-force-induced $C$ values, including those
taking $C_t^J = C_t^T \equiv 0$ will be called later
Skyrme-{\sc ledf} ({\sc s-ledf}).


\section{The spin fields}\label{sfield}

\begin{figure}
\begin{center}
\includegraphics[scale=0.70, angle=0.0, clip, viewport= 180 300 600
670]{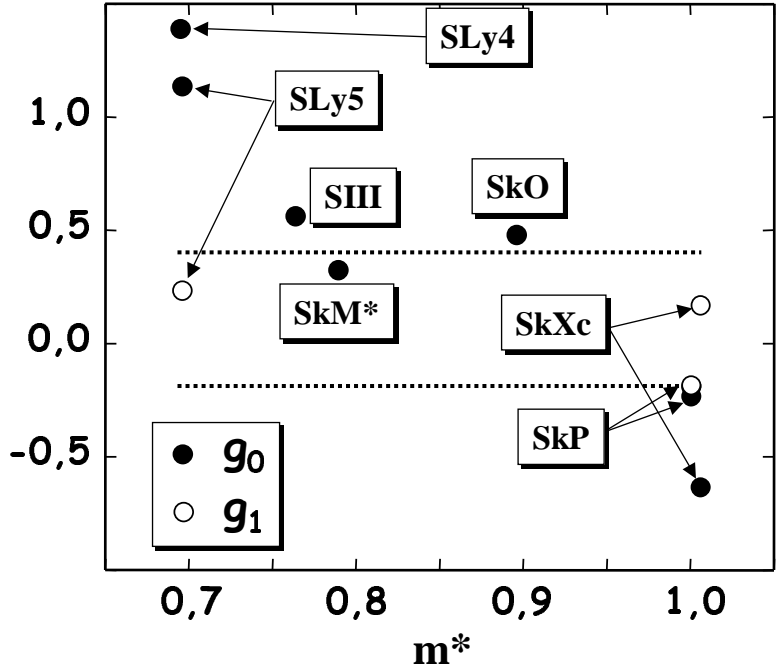}
\caption[]{Isoscalar Landau parameters $g_0$
(solid circles) and $g_1$ (open circles) for various
Skyrme forces. Vertical lines marks the values $g_0$=0.4 and $g_1$=--0.19
recommended by Bender~\etal~\protect{\cite{[Ben02]}}
from their study of the Gamow-Teller resonances.} \label{landau}
\end{center}
\end{figure}

The strengths of the spin fields, $C^s_t {\boldsymbol s}^2$
and $C^{\Delta s}_t {\boldsymbol s}$$\cdot$$\Delta{\boldsymbol s}$,
emerging from the Skyrme force appear to be  more or less accidental. This is
illustrated in Fig.~\ref{landau} showing
the isoscalar Landau parameters $g_0$ and $g_1$  calculated for the
Skyrme forces under study.  The Landau parameters
are related to the {\sc ledf} strengths in the following way~\cite{[Ben02]}:
\be\label{lanis}
g_0 = N_0 (2C_0^s + 2C_0^T \beta \rho_0^{2/3}), \quad \quad
g_1 = -2 N_0 C_0^T \beta \rho_0^{2/3},
\ee
\be\label{laniv}
g_0^{\,\prime} = N_0 (2C_1^s + 2C_1^T \beta \rho_0^{2/3}), \quad \quad
g_1^{\,\prime} = -2 N_0 C_1^T \beta \rho_0^{2/3},
\ee
where $\beta = (3\pi^2/2)^{2/3}$, and $N_0^{-1} = \pi^2 \hbar^2 / 2m^\star k_F$
is an effective-mass-dependent normalization factor.
The Skyrme-force-induced $g_0$ and $g_1$
parameters, see Fig.~\ref{landau}, are indeed scattered rather randomly
reflecting the fact that Skyrme forces are
fitted ultimately to the {\sc te} channel while the {\sc to} components
of the {\sc s-ledf} are only
cross-checked mostly through the high-spin (cranking) applications. In
Ref.~\cite{[Ben02]} the preferred values $g_0$=0.4,
$g_0^{\,\prime}$=1.2 and  $g_1$=--0.19, $g_1^{\,\prime}$=0.62
have been established from the analysis of the Gamow-Teller resonances,
see also Ref.~\cite{[Ost92]} and refs. quoted therein.  These
values were obtained under an additional assumption
of the density independence of $C_t^{s}$ and for $C_t^{\Delta s}$$\equiv$0.
The {\sc ledf} with spin fields defined in this way will be called
later the Landau-{\sc ledf} ({\sc l-ledf}).


\begin{figure}
\begin{center}
\includegraphics[scale=0.70, angle=0.0, clip, viewport= 20 40 550
800]{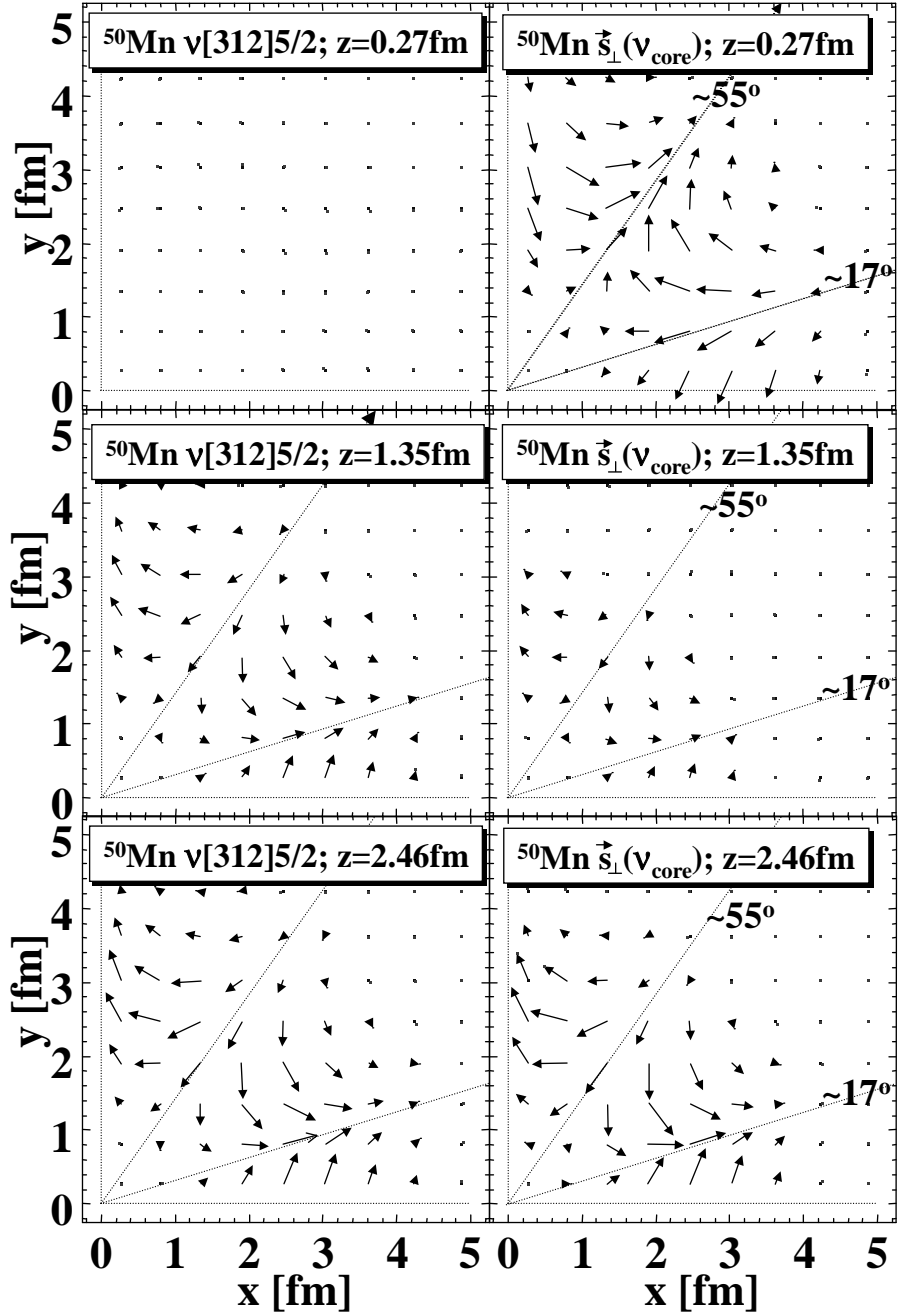}
\caption[]{The $\vec {\boldsymbol s}_{\perp}$ component
of the spin density (in arbitrary units) in the ground state of $^{50}$Mn,
calculated for three selected cross sections at $z$=0.27\,fm (upper part)
$z$=1.35\,fm (middle part), and $z$=2.46\,fm (lower part).
Left panels show the contribution of the odd neutron
right panels show polarization
effect exerted by the odd neutron on the neutron core.
For this case $\vec {\boldsymbol s}_{\perp}^{\,\nu} \approx
\vec {\boldsymbol s}_{\perp}^{\,\pi}$,
see text for further detail.}
\label{sxy}
\end{center}
\end{figure}

\subsection{The spin fields at the ground state}

Before proceeding to the study of the terminating
states let us discuss the spin fields of the ground states
(calculated in HF approximation) of $N$$\approx$$Z$ nuclei.
It is known that the
binding energies calculated using the complete {\sc s-ledf}
functional exhibit a rather peculiar behavior in
odd-odd $T_z=0$ and some odd-$A$ $|T_z|=1/2$ nuclei~\cite{[Sat99]}.
It manifests itself via additional binding energy
of the order of $\sim 1$\,MeV
as compared to the {\sc shf} calculations using only the {\sc te}
part of the {\sc s-ledf}. The effect disappears
for $|T_z| > 1$ nuclei.

\begin{figure}
\begin{center}
\includegraphics[scale=0.70, angle=0.0, clip, viewport= 60 370 800
650]{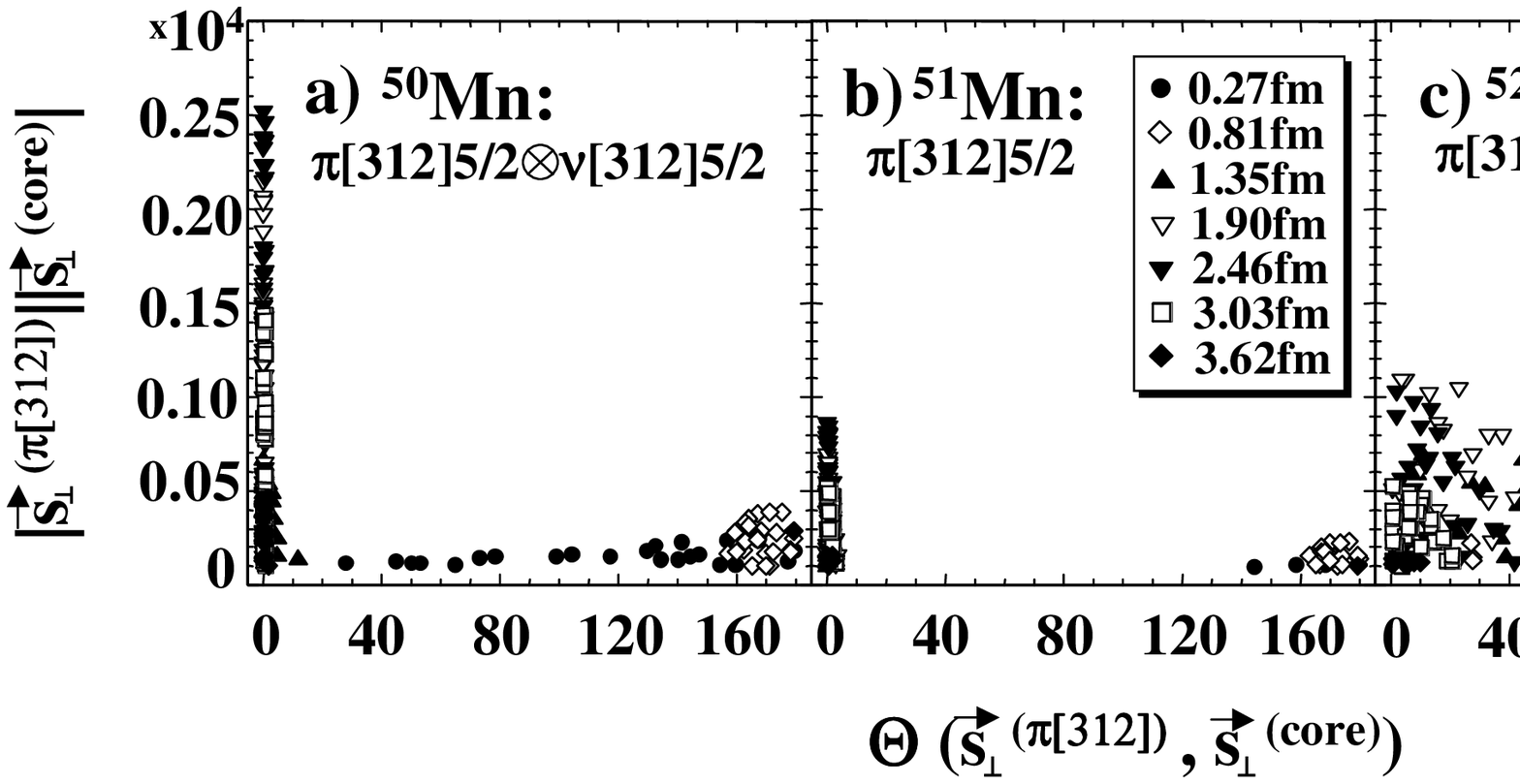}
\caption[]{Product of $|\vec s^{\,\pi[512]}_{\perp}({\boldsymbol r})|
|\vec s^{\,core}_{\perp}({\boldsymbol r})|$ reflecting the magnitude
of the spin fields (in arbitrary units) versus the classical
angle between  these vectors $\theta$, corresponding to their relative
orientation. {\bf a}) $N$=$Z$ o-o nucleus $^{50}$Mn,
{\bf b}) $N-Z=1$ odd-$A$ nucleus $^{51}$Mn, {\bf c}) $N-Z=2$ o-o nucleus
$^{52}$Mn.  All points calculated
at a fix value of the $z$ coordinate
and different values of the ($x$,$y$) coordinates are labeled
by the same symbol, as indicated in the legend.
} \label{sxycor}
\end{center}
\end{figure}

The enhancement in the binding energy of $N$$\approx$$Z$ nuclei
is due to a strong polarization effect exerted by  the spin field of the
odd-particle(s) on the spin field of the core, as illustrated in
Figs.~\ref{sxy} and \ref{sxycor} for a representative example of the
manganese isotopes.
Fig.~\ref{sxy} shows the $Oxy$ component of the spin density
$\vec{\boldsymbol s}_{\perp}$ in $^{50}$Mn for three selected
cross sections through
this axially deformed nucleus,  that include the
near-equatorial plane at $z$=0.27\,fm as well as $z$=1.35\,fm and
2.46\,fm planes. To visualize the polarization effect we
decompose the spin density into contributions of the valence particles (note
that $\vec{\boldsymbol s}^\pi \approx \vec{\boldsymbol s}^\nu$ in this $N$=$Z$
nucleus) and the core. The topology of the surfaces shown on the left hand side
reflects the structure of the dominant asymptotic [312]5/2 Nilsson
component in the wave
function of the valence particle.
Indeed, the one particle contribution to the spin
field in a simplex-conserving axial HO basis state:
\be
\Psi_{N n_z |\Lambda|; s=\pm i} =
\frac{1}{\sqrt2} ( \Psi_{N n_z \Lambda; 1/2} \pm i \Psi_{N n_z -\Lambda; -1/2}
) \quad \mbox{where} \quad  \Psi_{N n_z \Lambda} = \psi_{N n_z |\Lambda|}
e^{i\Lambda\varphi},
\ee
is
\be\label{sx-con}
s_x =  -\frac{1}{2} |\psi_{N n_z |\Lambda|} (\rho, z) |^2 \sin
2\Lambda\varphi,
 \quad
s_y =  -\frac{1}{2} |\psi_{N n_z |\Lambda|} (\rho, z) |^2 \cos
2\Lambda\varphi,
\quad
s_z = 0,
\ee
see~Ref.~\cite{[Dob00]} for further details.
Hence, $ |s_x|/|s_y| =|\mbox{ctg} 2\Lambda\varphi|$
and the plot of $\vec{\boldsymbol s}_\perp$ shows
the characteristic {\it "vortex"} lines for  $ |s_x|/|s_y| = 1$, i.e., for
$\varphi = [45^\circ + n\pi]/2\Lambda$  where $n=0,1,2,...$.
Two such lines that
appear in Fig.~\ref{sxy} are consistent with $\Lambda =2$.
Moreover, the small values of $s_{xy} \sim 0$ over the entire
equatorial plane at $z=0.27$\,fm are due to $\psi_{[312]} \sim H_1(z\approx 0)
\approx 0$. Calculations also show that although $s_z \ne 0$ the condition
$|s_z| \ll |s_{\perp}|$ is well fulfilled for most cases.

Correlation between the spin field
$\vec {\boldsymbol s}^{\,\pi[512]}_{\perp}({\boldsymbol r})$ due to
the occupation of  the [512]5/2 orbital by the
valence
proton and that of
the core $\vec {\boldsymbol s}^{\,core}_{\perp}({\boldsymbol r})$
(polarization effect) is illustrated in Fig.~\ref{sxycor}.
The figure shows  the product
$|\vec {\boldsymbol s}^{\,\pi[512]}_{\perp}({\boldsymbol r})|
\cdot |\vec {\boldsymbol s}^{\,core}_{\perp}({\boldsymbol r})|$ that
reflects the magnitude of the spin fields versus the classical angle between
these vectors $\theta$ that reflects their relative orientation.  The
figure clearly illustrates that in an o-o $N$=$Z$ nucleus
the core polarization
is strongest and of almost purely ferromagnetic type. In an odd-A
$N$$-$$Z$=$\pm$1 nuclei the effect is quenched but remains to be of
ferromagnetic type.
Hence,  the role of the spin fields in these nuclei is maximal. In $|N-Z| > 1$
nuclei the induced (core) spin field appears to be quenched
even further and, additionally, not coherent
with the valence-particle(s) spin field, particularly in o-o nuclei,
as depicted in Fig.~\ref{sxycor}c.

At least two very important conclusions can be drawn from this
analysis. ({\it i\/}) The coherence of the spin fields in
$N$$\sim$$Z$ nuclei may cause
a strong polarization of the nucleus. ({\it ii\/})
The magnitude of the spin-field-induced effects
is predicted to depend strongly on
isospin. These two observations give an unique opportunity
to  resolve the strength of the spin fields,
in particular, by using high spin states where
spin fields are expected to be enhanced.

\subsection{The spin fields at the band termination}\label{sf-bt}

\begin{figure}
\begin{center}
\includegraphics[scale=0.65, angle=0.0, clip, viewport= 30 290 800
650]{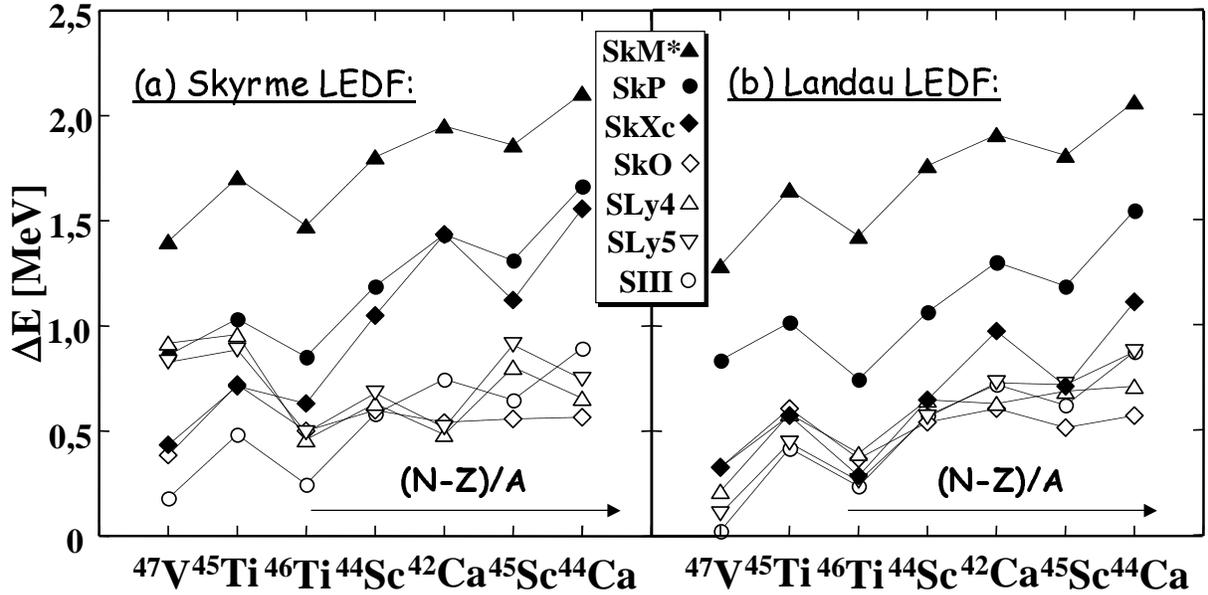}
\caption[]{Calculated
energy differences  for the terminating states
$\Delta E_{th} = E_{th}[d_{3/2}^{-1} f_{7/2}^{n+1}] -
E_{th}[f_{7/2}^{n}]$ relative to the experimental data $\Delta E \equiv \Delta
E_{exp} - \Delta E_{th}$ where $\Delta E_{exp}$ are listed in
Tab.~\protect{\ref{edata}}. The left panel shows \shf~calculations while right
part illustrates calculations using {\sc l-ledf}.}
\label{deltaE}
\end{center}
\end{figure}

Fig.~\ref{deltaE}a shows the calculated
energy differences  for the terminating states
$\Delta E_{th} = E_{th}[d_{3/2}^{-1} f_{7/2}^{n+1}] -
E_{th}[f_{7/2}^{n}]$ relative to the experimental data $\Delta E_{exp}$ given
in Tab.~\ref{edata}, i.e. the values of $\Delta E \equiv \Delta E_{exp} -
\Delta E_{th}$.  The first striking observation stemming from this calculation
is that {\it all\/} considered Skyrme forces systematically underestimate
empirical data by at least 10\% .
In case of the SkM$^*$ and SkP parameterizations the
difference even exceeds 20$\div$30\%.
The disagreement is unexpectedly
large provided the structural simplicity of the terminating states.

Let us further observe that the values of $\Delta E$
calculated using SLy4 and SLy5 forces, which are in general
rather similar to $\Delta E$  obtained using SIII or SkO forces,
increase rapidly in $N$$-$$Z$=1 nuclei $^{47}$V and
$^{45}$Ti. This result is related to the magnitude of the spin fields
and the ferromagnetic-type polarization of the core which, as discussed
in the preceding section and in Ref.~\cite{[Sat99]},
are exceptionally large for Lyon-forces, see also Fig.~\ref{landau}.
Since no enhancement of this type is observed in the data,
this result clearly shows that an unified description
of the spin fields
within the {\sc ledf} theory is required.

Hence, we state that the
experimental data suggests a generalization of the  {\sc s-ledf}.
However, our strategy is to introduce a  minimal-type
modification that pertains ultimately to the {\sc to} part
of the {\sc s-ledf} only. More precisely, for SLy4, SIII, SkO, and SkM$^*$
we will change the spin
fields i.e. the first two terms of the {\sc to} part of the {\sc ledf}
(\ref{todd}) not affecting the local gauge invariance (\ref{gauge}).
For SkP, SkXc, and SLy5 on the other hand, we also slightly modify the {\sc to}
part of the tensor term. In this way
we actually brake the local gauge invariance.
However, our calculations show that this has a very small effect
on the final results
when compared to calculations using the
Skyrme-force-induced $C_t^T$ values. Thus, one can state that
the local gauge invariance (\ref{gauge}) is in fact preserved
in our calculations.

Our favorite unification scheme for the treatment of spin fields
(called {\sc l-ledf}) follows the one developed by
Bender~\etal~\cite{[Ben02]}. Let us recall, that
in the {\sc l-ledf} calculations we assume density
independent coefficients $C_t^s$ defined through
the Landau parameters $g_0$=0.4,
$g_0^{\,\prime}$=1.2, $g_1$=--0.19, and $g_1^{\,\prime}$=0.62 and set
$C_t^{\Delta s}$=0.  Such a simple treatment of the spin fields leads to a
surprisingly consistent picture for various Skyrme-forces, irrespectively
the difference in effective mass.
 Indeed, the results obtained for all forces, except of
SkM$^*$ and SkP, essentially overly each other as shown in
Fig.~\ref{deltaE}b. Let us further observe that $\Delta E$ calculated
with SkM$^*$ and SkP  show almost a constant offset as compared to
the other forces.   We suspect that
the effective mass or, equivalently, current independence of our results is
directly related to the gauge invariance of the {\sc ledf}. This point
requires, however, further investigation.

\begin{figure}
\begin{center}
\includegraphics[scale=0.65, angle=0.0, clip, viewport= 30 260 800
800]{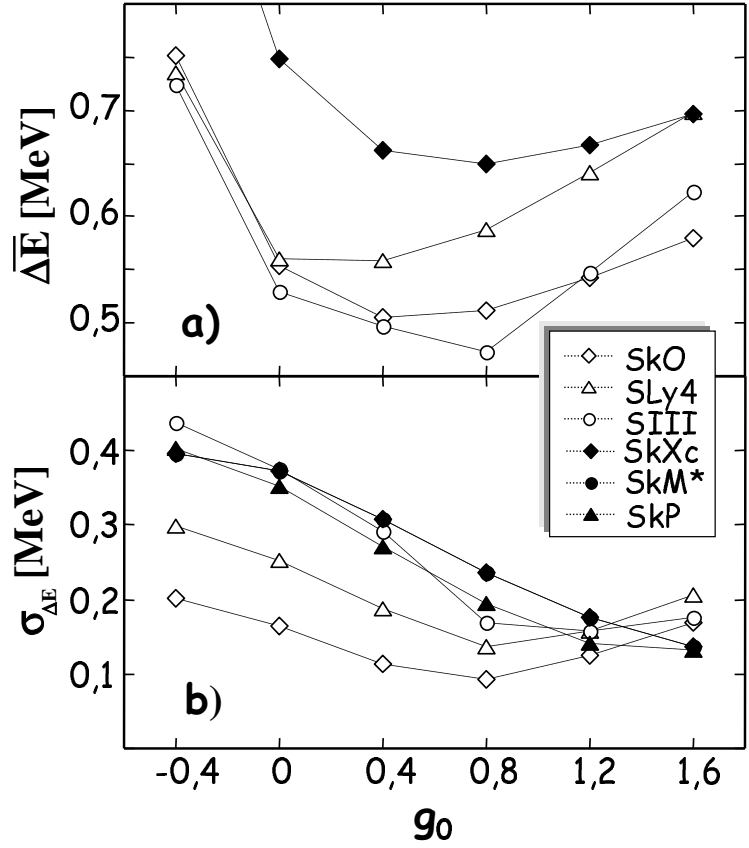}
\caption[]{The average difference  $\overline{\Delta E}$
($\Delta E \equiv \Delta E_{exp} - \Delta E_{th}$) (upper part)
and standard deviation $\sigma$ (lower part) between the data
and the calculations versus Landau parameter $g_0$.
}
\label{averg0}
\end{center}
\end{figure}

Since we do concentrate on $N$$\sim$$Z$ nuclei, our calculations are
essentially insensitive to changes in the isovector Landau parameters
$g_0^\prime$ and $g_1^\prime$ in the wide range of their proposed values.
Sensitivity of our predictions with respect to the isoscalar Landau parameter
$g_0$ is illustrated in Fig.~\ref{averg0}.
Fig.~\ref{averg0}a shows the average deviation from the data,
$\overline{\Delta E}$, versus $g_0$.
Note, that $\overline{\Delta E}$ is minimal for
$g_0 \sim 0.4\div 0.8$, what is very close to the suggested value
$g_0$=0.4 of Ref.~\cite{[Ben02]}.
Let us further observe that $\overline{\Delta E}$
does not change sharply within the interval
$\Delta g_0 = \pm 0.4$ around the preferred value, but
our analysis seem to rule out both negative
and large positive $g_0 > 1.2$ values of $g_0$.
Moreover, it clearly shows that the $\sim$10\%
discrepancy  between the calculations and the data cannot be
accounted for by further readjusting the Landau parameters,
at least not within the analyzed unification scheme.

Fig.~\ref{averg0}b shows the dependence of the standard deviation,
$\sigma_{\Delta E}$, which reflects the spread in
$\Delta E$, on $g_0$. Apparently the minimum is obtained
for $g_0 \sim 0.8\div 1.2$, i.e. well above the preferred
value of $g_0$=0.4. Let us observe, however,
that almost all curves shown in Fig.~\ref{deltaE}
show clearly an increasing trend as a function of the
reduced isospin $T_A\equiv (N-Z)/A$. Hence, part of
the spread may merely reflect the isovector properties of the
{\sc ledf}, most likely, the isovector part of the \ls-term.
The relatively weak dependence of $\Delta E$ on $T_A$ obtained
for the SkO force seems to support this conclusion.
Additional analysis strengthening this scenario will be given in
the next section.

\section{The spin-orbit term}\label{so}

\begin{figure}
\begin{center}
\includegraphics[scale=0.7, angle=0.0, clip, viewport= 150 270 650
800]{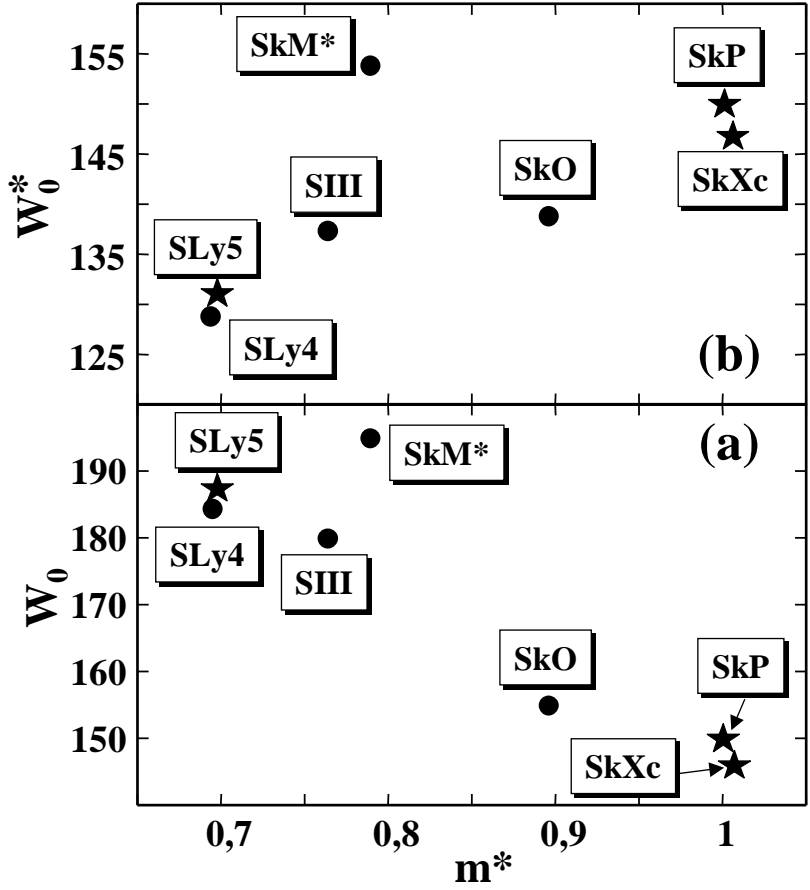}
\caption[]{The strength $W_0$ (lower part) and the effective-mass scaled
strength $W_0^\star$ (upper part) of the isoscalar part of the spin-orbit \shf~
potential versus the effective mass. Filled dots denote parameterizations
which do not include tensor densities while those including these terms
are indicated by stars. See text for more detail.}
\label{vso}
\end{center}
\end{figure}

Within the {\sc shf} theory, the
$\ell s$-potential takes the following form:
\be\label{so1}
V_{LS}(q, \boldsymbol{r}) = - i \boldsymbol{W}_q (\boldsymbol{r})
\boldsymbol{\nabla}\times \boldsymbol{\sigma},
\ee
where
\be\label{so2}
\boldsymbol{W}_q (\boldsymbol{r}) =
\frac{1}{2}  W {\boldsymbol \nabla}\rho_0 (\boldsymbol{r})
+ \frac{1}{2} W^{\,\prime} {\boldsymbol \nabla}\rho_q (\boldsymbol{r})
- \frac{1}{8} \left[ (t_1 x_1 + t_2 x_2) \boldsymbol{J}(\boldsymbol{r})
   + (t_2 - t_1) \boldsymbol{J}_q (\boldsymbol{r}) \right].
\ee
The vector spin density $\boldsymbol{J}(\boldsymbol{r})$ dependent terms appear
to contribute rather weakly to the $\ell s$-potential. Hence, the magnitude
of the $\ell s$-potential is determined essentially by the first two terms in
Eq.~(\ref{so2}). For the spherical limit the $\ell s$-potential can be
approximated by:
\be\label{so3}
V_{LS}(q, r)  \approx
\left\{  W \frac{1}{r} \rho_0^{\,\prime} (r)
+ W^{\,\prime} \frac{1}{r} \rho_q^{\,\prime} (r) \right\} {\boldsymbol \ell}
{\boldsymbol s} =
\left\{  \frac{W_0}{r} \rho_0^{\,\prime} (r) \pm
 \frac{W_1}{r} \rho_1^{\,\prime} (r)
\right\} {\boldsymbol \ell} {\boldsymbol s}
\ee
where
$\rho_0^{\,\prime} = (\rho_n + \rho_p)^{\,\prime}$
and $\rho_1^{\,\prime} = (\rho_n - \rho_p)^{\,\prime}$
are the radial derivatives of
the local isoscalar and isovector densities, while
$W_0 \equiv W + \frac{1}{2} W^{\,\prime}$
and $W_1 \equiv \frac{1}{2} W^{\,\prime}$ denote
the isoscalar and isovector
strengths, respectively.

A direct comparison of the isoscalar strengths $W_0$ of the
\ls-potential is given in Fig.~\ref{vso}a. Apparently,
SLy4, Sly5, SkM$^*$, and SIII forces have strong, while SkO, SkP,
and SkXc have a weak \ls-potential assuming, of course, that there are no
drastic differences in the isoscalar density profile.
The latter assumption should
be rather well fulfilled in light nuclei,
which are considered here. It is interesting to note that the conclusions
stemming from a direct comparison of $W_0$ are in complete contradiction to the
results presented in Fig.~\ref{deltaE}. Indeed, according to our calculations
Sly4, SLy5, SkO, SIII, and eventually also SkXc are expected to have similar
\ls-strength while it should be
considerably stronger for SkM$^*$ and SkP.

The question therefore arises of how to compare the
strengths of the \ls-potential for
different parameterizations. The problem appears to be related to non-local
effects which are, within the {\sc shf}, absorbed into the kinetic
energy term through the effective mass $m^*$.  The impact
of non-localities on the \ls-potential can be studied
using  the so called asymptotically equivalent
wave function~\cite{[Bei75],[Lop00]}
\be\label{asymwf} \tilde \phi_i
(\boldsymbol{r}) =   \sqrt{ \frac{ m }{ m^\star (\boldsymbol{r}) }} \phi_i
(\boldsymbol{r}).
\ee
This representation allows to
rewrite the \shf~equations  in an alternative form
with a bare mass $m$ in the kinetic energy
term, a state-dependent central  potential $U_q (e_\mu;\boldsymbol{r})$, and
the effective-mass-scaled \ls-potential (\ref{so3}):
\be\label{hfso}
V_{LS}(q,r)
   \approx \frac{ m^\star (r) }{ m }
\left\{  \frac{W_0}{r} \rho_0^{\,\prime} (r) \pm
  \frac{W_1}{r} \rho_1^{\,\prime} (r)
\right\} {\boldsymbol \ell} {\boldsymbol s}.
\ee
The effective-mass-scaled isoscalar strengths
$W_0^*\equiv \frac{m^\star}{m}W_0$ are depicted in
Fig.~\ref{vso}b. Note, that the classification of the \ls-strength according to
$W_0^*$ agrees very nicely with our results shown in Fig.~\ref{deltaE}.
Indeed, the values of $W_0^*$ are similar for Sly4, SLy5, SIII, and SkO and
considerably  larger for SkP and SkM$^*$.

This indicates that at least part of the observed
discrepancy, $\overline{\Delta E}$, results from a too strong \ls-term.
By reducing the strength with
 $\sim 5$\%, the $\overline{\Delta E}$ reduces by
$\sim$350\,keV bringing it to an acceptable level of
$\overline{\Delta E}\sim 200$\,keV for most of the forces.
In particular, for the case of the SkO force
$\overline{\Delta E}$ drops from 504\,keV to 164\,keV, while for Sly4
a 5\% reduction of the \ls-strength decreases
$\overline{\Delta E}$  from 560\,keV to 189\,keV.

\vspace{0.3cm}

Concerning the isovector \ls-potential, the Skyrme
forces which are discussed in the literature
can be divided into three major classes.
The standard Skyrme force parameterizations
assume that $W=W^{\,\prime}$
($W_1/W_0=1/3$), implying that ${\boldsymbol W}_q \sim
W (2\rho_q^{\,\prime} + \rho_{-q}^{\,\prime})$.
The Sly4, Sly5, SkM$^*$, SkO, and SIII are standard forces
among those studied here.

Non-standard Skyrme interactions with $W\ne W^{\,\prime}$ were first
studied by Reinhard and Flocard~\cite{[Rei95]} in connection
with isotope shifts in Pb nuclei. Consistency with experimental
data led them to the parameterizations with $W^{\,\prime} \sim - W$
or ($W_1/W_0=-1$),
i.e. to an entirely different isovector dependence of the
$\ell s$-term ${\boldsymbol W}_q \sim
W \rho_{-q}^{\,\prime}$ as compared to the standard one.
The study of the \ls-term in neutron-rich nuclei
by Reinhardt~\etal~\cite{[Rei99]} seems further
to corroborate this result. The so called SkO
parametrization  established in Ref.~\cite{[Rei99]} (and studied here)
have even larger negative value of $W_1$ with $W_1/W_0 \approx -1.3$.

The third type of the \ls-term which is considered in the literature
in connection with the {\sc shf} approach, was introduced
by Brown~\cite{[Bro98]} who uses $W^{\,\prime}=0$ (parametrization SkXc). In
this case there is no isovector \ls-term ($W_1/W_0\equiv 0$) and
${\boldsymbol W}_q \sim W (\rho_{n}^{\,\prime}
+ \rho_{p}^{\,\prime})$.

\begin{figure}
\begin{center}
\includegraphics[scale=0.7, angle=0.0, clip, viewport= 150 270 650
800]{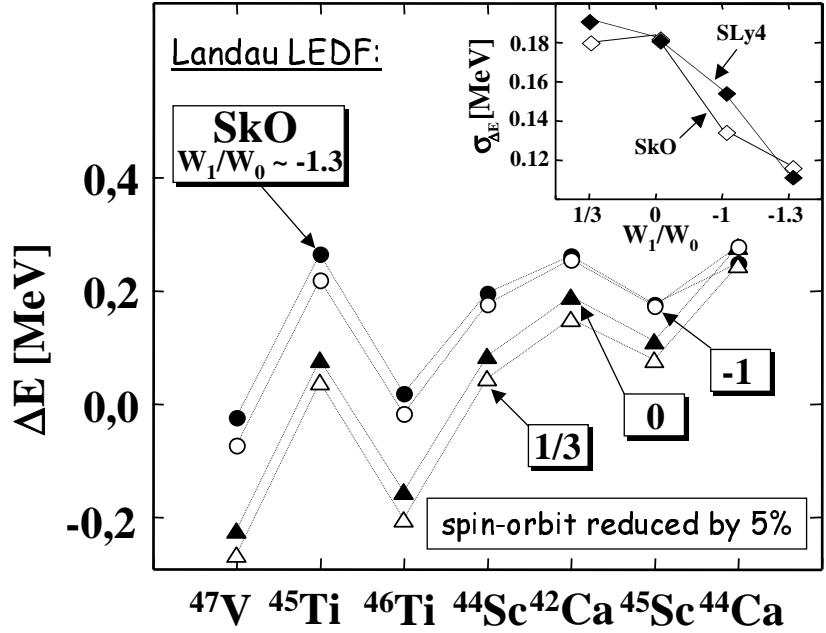}
\caption[]{The values of $\Delta E$ calculated using
SkO-induced {\sc l-ledf} with four different parameterizations of the
isovector \ls-term, including the original $W_1/W_0\approx -1.3$ strength
and the modified strengths $W_1/W_0= -1,0,1/3$. In the calculations
$W_0$ was kept fix at 5\%
below its original value. The insert shows
the dependence of the spread in  $\Delta E$, $\sigma_{\Delta E}$,
on  $W_1/W_0$ calculated for SkO-induced as well as SLy4-induced
{\sc l-ledf}.
}
\label{skoso}
\end{center}
\end{figure}

\vspace{0.3cm}

As already discussed at the end of Sect.~\ref{sf-bt}
our calculations give certain preference for the SkO-induced
{\sc l-ledf}, since
it minimizes the spread in $\Delta E$, $\sigma_{\Delta E}$.
In particular, this result seems to speak in favor
of a \ls-potential with large negative isovector strength
$W_1$. To reinforce this observation we have performed a set of
calculations based on SkO-induced {\sc l-ledf}, but
exploring different isovector dependence of
the \ls-term, including the four possibilities
discussed above $W_1/W_0 = -1.3, -1, 0, 1/3$.  In the
calculations, the isoscalar strength
$W_0$ was kept constant and its value was reduced by 5\%
as compared to the original SkO strength.


The calculated values of $\Delta E$ for these four variations of the
SkO-induced {\sc l-ledf} are shown in Fig.~\ref{skoso}.
Drastic change in the $W_1/W_0$ ratio from the (near)original
values --1.3,\,--1 to 0,\,1/3 clearly destroy the agreement
to the data.  Indeed, the spread, $\sigma_{\Delta E}$,
increases from  113\,keV and 136\,keV ($W_1/W_0= -1.3,-1$) to
184\,keV and 180\,keV ($W_1/W_0= 1/3,\, 0$), respectively, see
insert in Fig.~\ref{skoso}. Note, however, that since
we deal with $N$$\sim$$Z$, a fine tuning of the isovector
terms cannot be achieved.

Similar calculations, but with the SLy4-induced
{\sc l-ledf} additionally corroborate our conclusions.
The change in $W_1/W_0$ ratio from the (near)original
values 1/3,\,0 to --1,\,--1.3 improve the agreement
to the data, as shown in the insert in Fig.~\ref{skoso}.
Note also, that the calculated spread, $\sigma_{\Delta E}$,
is quantitatively very similar for both the SkO-induced and
SLy4-induced {\sc l-ledf} provided the isovector part of
the \ls-term is similar.

\section{$2p$-$2h$ terminating state in $^{45}$$\mbox{Sc}$ and the additivity
of $1p$-$1h$ excitations.}\label{addit}

\begin{figure}
\begin{center}
\includegraphics[scale=0.7, angle=0.0, clip, viewport= 150 270 650
800]{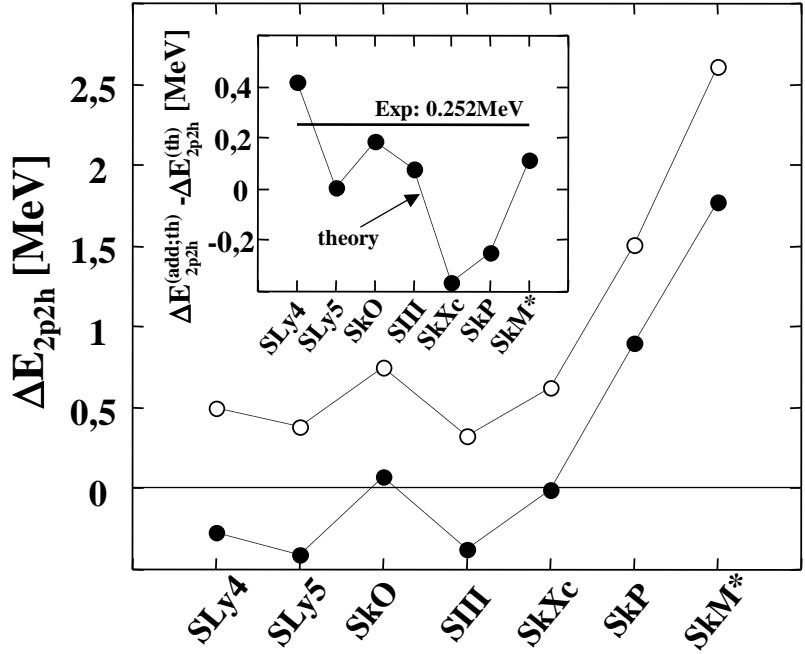}
\caption[]{$\Delta E_{2p2h}$ calculated for different
Skyrme-force-induced {\sc l-ledf}. Open circles mark calculations
using the standard \ls~strength, while  those using a 5\%
reduced \ls~strength are labeled by filled circles. The insert shows
the theoretically calculated deviation from the additivity
$\Delta E_{2p2h}^{(add;th)}-\Delta E_{2p2h}^{(th)}$.
The corresponding experimental value of
$\Delta E_{2p2h}^{(add;exp)}-\Delta E_{2p2h}^{(exp)}$=0.252\,MeV is
marked by a solid line. }
\label{2p2h}
\end{center}
\end{figure}

The $2p$-$2h$ $d_{3/2}^{-2}f_{7/2}^{n+2}$ terminating states
can provide further insight into the \todd~terms,
the \ls~potential, and into various polarization phenomena,
since they allow to test
self-consistent calculations against simple, intuitive
additivity relations.  Thus far such state is known, however, only in
$^{45}$Sc~\cite{[Lenzi]}. Although a single case does not allow for deep
conclusions, the comparison between the data and the calculations, see
Fig.~\ref{2p2h}, clearly shows that some {\it naively-understood\/}
additivity concepts do not always work.

For example, the values of $\Delta E_{2p2h}\equiv \Delta E_{2p2h}^{(exp)}
- \Delta E_{2p2h}^{(th)}$ calculated using Sly4,
SLy5, SIII, and SkXc {\sc l-ledf} are comparable or even smaller than
the corresponding values of $\Delta E$ obtained for the
terminating $1p$-$1h$ state
in $^{45}$Sc, see Fig.~\ref{deltaE}. Consequently, calculations using a 5\%
reduced isoscalar \ls-strength overestimate the data with
exception of the SkXc {\sc l-ledf} for which one obtains an almost perfect
agreement to the data.
On the contrary, the additivity seems to work perfectly for the SkO-induced
{\sc l-ledf}. A reduction of the \ls-strength by 5\%
gives, in this case, an excellent agreement to the data.

The excitation energy of the $2p$-$2h$
terminating state in $^{45}$Sc
can be approximated by a sum of the energy of the two aligned $29/2^+$,
$[d_{3/2}^{-1}f_{7/2}^{n+1}]_{(I_{max}-1)}$ states terminating at
unfavored signature. Since there exist exactly two
such states, their centroid energy (sum of their energies) should be
fairly insensitive to the possible beyond-mean-field mixing effects.
Hence, the additivity relation for the excitation energies should
provide an independent characteristics of the {\sc ledf}.

Insert in Fig.~\ref{2p2h}
shows the theoretically calculated difference $\Delta E_{2p2h}^{(add;th)} -
\Delta E_{2p2h}^{(th)}$ where $\Delta E_{2p2h}^{(th)}$ denotes
excitation energy of the terminating $2p$-$2h$ state while $\Delta
E_{2p2h}^{(add;th)}$ denotes $2p$-$2h$ excitation energy
calculated by adding excitation energies of the two $1p$-$1h$
configurations given above.
In general, the additivity works relatively well. There are
however two exceptions: the SkP and,
unexpectedly, SkXc {\sc l-ledf}.
The downward shift of SLy5 with respect to SLy4 as well
as rather mediocre agreement for SkP and SkXc forces, may be is related to the
tensor component that is present in these three forces and deserves further
investigation.

This short analysis indicates already the rich physics that can be addressed
using $2p$-$2h$ terminating states.

\section{Summary}

\begin{figure}
\begin{center}
\includegraphics[scale=0.7, angle=0.0, clip, viewport= 150 270 650
800]{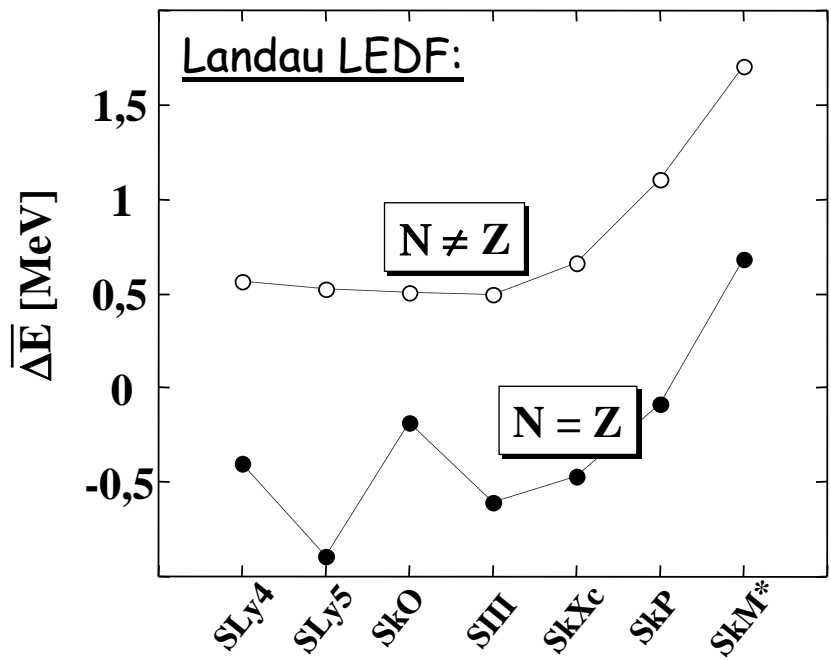}
\caption[]{The average deviation $\overline{\Delta E}$  for $N$$\ne$$Z$
(open dots) and for $N$$=$$Z$ (filled dots) nuclei. The latter case includes
data in $^{44}$Ti~\protect{\cite{[Lenzi]}}
where $E[f_{7/2}^{n}]_{12^+}$=8.038\,MeV,
$E[d_{3/2}^{-1}f_{7/2}^{n+1}]_{15^-}$=13.369\,MeV, and
$\Delta E_{exp}$=5.331\,MeV,
and $^{46}$V~\protect{\cite{[Len99]}}
where $E[f_{7/2}^{n}]_{15^+}$=8.484\,MeV,
$E[d_{3/2}^{-1}f_{7/2}^{n+1}]_{17^-}$=13.629\,MeV, and
$\Delta E_{exp}$=5.145\,MeV.
All calculations were done using
{\sc l-ledf}. }
\label{neqz}
\end{center}
\end{figure}

We have performed a systematic study of terminating states in the $A$$\sim$50
mass region using the self-consistent Skyrme-Hartree-Fock
model and testing several parameterizations of the Skyrme force.
The objective was to demonstrate that
the terminating states, due to their intrinsic simplicity, offer
an unique and so far unexplored opportunity to study different
aspects of effective NN interaction
or nuclear local energy density functional
within the self-consistent approaches.

It is shown, that
the Skyrme-force parameterizations used in our work,
including Sly4, Sly5, SkO, SIII, SkXc, SkP, and SkM$^*$, give rise to
a rather mediocre (for SkP and SkM$^*$ even unacceptable) agreement with
the data even for such a seemingly simple observable like the energy
difference between the aligned $f_{7/2}^{n}$ and
$d_{3/2}^{-1}f_{7/2}^{n+1}$ states.

It is further demonstrated that a simple unification of
the spin fields according to the scheme proposed in Ref.~\cite{[Ben02]}
leads to an unified description of the data for
Sly4, Sly5, SkO, SIII, SkXc, i.e. for very different
parameterizations. This result seems to indicate the importance of
the local gauge (and Galilean) invariance
of the local energy density functional.
The remaining discrepancy of $\sim$500\,keV
(see Fig.~\ref{neqz}) which is still of the order of $\sim$10\%
cannot be reduced by further readjusting the spin fields.

It is also shown, that the observed disagreement between theory
and experiment for different parameterizations correlates nicely
with the values of the effective mass scaled isoscalar
strength of the \ls-term for these parameterizations. Hence,
a part of this discrepancy can, most likely, be ascribed to a
too strong isoscalar \ls-term. Reduction of the
isoscalar \ls-strength  by 5\%
reduces the discrepancy well below the 5\% level.
Moreover, our calculations suggest that the spread in $\Delta E$
can be further reduced by adopting
the non-standard parameterizations of the \ls-term
with a strong negative isovector strength $W_1/W_0 \leq -1$.

Finally, let us point out that there is a large difference
in $\overline{\Delta E}$
calculated  in $N$$\ne$$Z$ and  $N$$=$$Z$
nuclei, see Fig.~\ref{neqz}. For $N$$\ne$$Z$ our {\sc l-ledf}~approach
systematically underestimates the data  while the opposite is true
for $N$$=$$Z$ nuclei. The offset between the two curves is of the order
of $\sim$1\,MeV. This result leaves us with an extremely important
question: Does this offset
indicate a
breakdown of the standard mean-field in  $N$$\sim$$Z$ nuclei
and a need for substantial configuration mixing even at
the terminating states~\cite{[Ter98]} and what is
the possible source of
such mixing?

\bigskip

\bigskip

This work was supported by the Foundation for Polish Science (FNP), the
G\"oran Gustafsson Foundation,
the Swedish Science Council (VR), the Swedish  Institute (SI) and
the Polish Committee for Scientific Research (KBN).
The authors thank Sylvia Lenzi and Jan Stycze\'n for communicating
experimental results prior to publication.
One of us (WS) is grateful to J. Dobaczewski for
stimulating discussions and many constructive comments on the manuscript.


\end{document}